\documentclass[pdflatex,iicol]{sn-jnl}

\usepackage[numbers,sort&compress]{natbib}
\usepackage{graphicx}
\usepackage{xcolor}
\usepackage{amsmath,amssymb}
\usepackage{placeins}
\usepackage{stfloats}
\usepackage{microtype}
\usepackage{xurl}
\hypersetup{
  hidelinks,
  pdftitle={Primordial B-mode sensitivity forecasts for northern-sky cosmic microwave background surveys},
  pdfauthor={Dongdong Zhang, Bo Wang, Jia-Rui Li, Yi-Fu Cai, Chang Feng}
}

\begin{document}

\title[Northern-sky B-mode sensitivity]{Primordial B-mode sensitivity forecasts for northern-sky cosmic microwave background surveys}

\author[1,2,3,4]{\fnm{Dongdong} \sur{Zhang}}
\author[1,3,4]{\fnm{Bo} \sur{Wang}}
\author[1,3,4]{\fnm{Jia-Rui} \sur{Li}}
\author*[1,3,4]{\fnm{Yi-Fu} \sur{Cai}}\email{yifucai@ustc.edu.cn}
\author*[1,3,4]{\fnm{Chang} \sur{Feng}}\email{changfeng@ustc.edu.cn}

\affil[1]{\orgdiv{Department of Astronomy, School of Physical Sciences}, \orgname{University of Science and Technology of China}, \orgaddress{\city{Hefei}, \state{Anhui}, \postcode{230026}, \country{China}}}
\affil[2]{\orgdiv{School of Aerospace Information}, \orgname{Hefei Institute of Technology}, \orgaddress{\city{Hefei}, \state{Anhui}, \postcode{238076}, \country{China}}}
\affil[3]{\orgdiv{School of Astronomy and Space Science}, \orgname{University of Science and Technology of China}, \orgaddress{\city{Hefei}, \state{Anhui}, \postcode{230026}, \country{China}}}
\affil[4]{\orgdiv{CAS Key Laboratory for Researches in Galaxies and Cosmology}, \orgname{University of Science and Technology of China}, \orgaddress{\city{Hefei}, \state{Anhui}, \postcode{230026}, \country{China}}}


\abstract{Northern-sky cosmic microwave background (CMB) polarization observations offer additional sky coverage and independent measurements in the search for primordial gravitational waves. We assess their potential through idealized sensitivity forecasts motivated by the Ali CMB Polarization Telescope. Using a power-spectrum likelihood and Markov chain Monte Carlo sampling, we relate assumed map depths, accessible angular scales, and effective sky coverage to constraints on the tensor-to-scalar ratio $r$. The model retains lensing B modes and assumes negligible residual foreground power. A benchmark with 95 and 150\,GHz channels, a polarization map depth of $40\,\mu\mathrm{K\,arcmin}$ per channel, and $f_{\rm sky}=0.1$ yields $\sigma(r)=0.023$ for a fiducial $r=0.014$, improving to 0.015 with Planck-like mock data. At a per-channel depth of $10\,\mu\mathrm{K\,arcmin}$, the northern survey alone reaches $\sigma(r)=0.0031$. Raising the minimum multipole from 30 to 70 degrades the baseline sensitivity to 0.037, while extending the effective area at fixed depth improves the constraint. As the northern maps deepen, their contribution increasingly determines the precision on $r$, whereas Planck-like information continues to constrain the scalar spectral index. These forecasts quantify the scope for improved northern-sky tensor constraints and establish sensitivity benchmarks for subsequent studies of joint northern and southern CMB observations.}


\maketitle

\section{Introduction}
\label{sec:introduction}

Inflation provides a framework for the early accelerated expansion of the universe and for the origin of primordial perturbations \citep{1980PhLB...91...99S,kazanas1980dynamics,sato1981first,1981PhRvD..23..347G,1982PhLB..108..389L,1982PhRvL..48.1220A,1983PhLB..129..177L}. Scalar fluctuations seed the cosmic microwave background (CMB) anisotropies and the growth of structure \citep{1981ZhPmR..33..549M,1982PhLB..115..295H,1982PhRvL..49.1110G,1982PhLB..117..175S,Mukhanov:1985rz,1986ApJ...304...15B}. Tensor fluctuations produce a B-mode polarization signal whose amplitude is conventionally described by the tensor-to-scalar ratio $r$ \citep{1997PhRvL..78.2054S,1997PhRvL..78.2058K,Hu:1997hv}. Measuring this signal would probe an energy scale and a sector of inflationary physics that are only indirectly constrained by the scalar perturbations.

The \textit{Planck} mission established precise constraints on the standard cosmological model and on inflation \citep{Planck:2018vyg,Planck:2018jri}. Ground-based instruments have pursued complementary measurements with increasing sensitivity and angular resolution \citep{Carlstrom:2009um,POLARBEAR:2015ixw,ACTPol:2016kmo,Dahal:2019xuf}. The BICEP/\textit{Keck} analysis of observations through the 2018 season, combined with \textit{Planck} and WMAP, obtained $r<0.036$ at 95\% confidence \citep{BICEP:2021xfz}, extending the constraints from data through the 2015 season \citep{BICEP2:2018kqh}. The small amplitude of the expected signal makes sensitivity, foreground control, and independent confirmation central to the search \citep{Kamionkowski:2015yta}. Published survey forecasts, including the CMB-S4 design study and the Simons Observatory science forecasts, provide quantitative treatments of these requirements \citep{CMB-S4:2020lpa,SimonsObservatory:2018koc}.

Northern observations can broaden the sky coverage of the international CMB program and provide independent measurements of a common cosmological signal. The scientific case for the Ali CMB Polarization Telescope (AliCPT) highlights access to northern low-foreground regions and complementarity with observations from Chile and the South Pole \citep{2016SCPMA..59g.178C,Li:2017lat,Li:2017drr,2018NatAs...2..104L}. In particular, the AliCPT science study discusses complementary coverage with BICEP and the Simons Array \citep{Li:2017drr}. Additional regions supply new CMB modes, while overlapping observations enable cross-spectra and checks across instruments and foreground environments. Atmospheric studies based in part on MERRA-2 support the high Tibetan Plateau as a site for millimeter-wave observations \citep{bosilovich2015merra,2018NatAs...2..104L}; GreenPol provides another northern observing concept \citep{Fuskeland:2023vof}.

AliCPT simulation studies have investigated foreground removal and the recovery of $r$ with several analysis pipelines \citep{Ghosh:2022mje,Dou:2023vqg,Zhang:2024fg}. In particular, \citet{Ghosh:2022mje} found sensitivities of order $\sigma(r)\sim0.02$ for the configurations considered. The data-challenge analyses also examine foreground residuals, sky selection, and consistency among cleaning methods \citep{Zhang:2024fg}. These results provide a reference for exploring the scope for further improvement: what precision on $r$ would deeper northern polarization maps deliver, how does that precision depend on the large-scale modes retained, and how does the role of external CMB information change as the maps deepen?

We address these questions with an idealized survey model and Markov chain Monte Carlo (MCMC) forecasts. Published AliCPT design and pipeline studies motivate a generic northern-hemisphere telescope benchmark, denoted NHT-1. We vary its map depth, minimum multipole, and effective sky fraction, and examine its combination with Planck-like mock data. Map depth is an input to the forecast, allowing each assumed sensitivity to be associated with a resulting precision on $r$. This establishes a quantitative starting point for assessing how improved northern observations could contribute to a global CMB program. A subsequent joint survey analysis can build on these benchmarks by incorporating southern observations, their overlap, and detailed noise, foreground, and instrumental models. Section~\ref{sec:framework} defines the survey model and likelihood, Section~\ref{sec:results} presents the forecasts, and Sections~\ref{sec:discussion} and \ref{sec:conclusions} discuss their implications.

\section{Survey Model and Forecasting Method}
\label{sec:framework}

\subsection{Benchmark Survey Configuration}
\label{sec:configuration}

Table~\ref{tab:survey} specifies the NHT-1 benchmark. The 95 and 150\,GHz bands and the beam widths are drawn from published AliCPT design studies \citep{Li:2017lat,Li:2017drr}. The per-channel map depth is a survey-level input whose variation describes possible future sensitivity. NHT-1 denotes this AliCPT-motivated forecast configuration.

\begin{table}[!htbp]
\centering
\caption{NHT-1 Benchmark Parameters}
\label{tab:survey}
\footnotesize
\begin{tabular}{cccc}
\hline
Frequency & $\Delta_{\rm P}$ & $\theta_{\rm FWHM}$ & $f_{\rm sky}$ \\
(GHz) & ($\mu$K arcmin) & (arcmin) & \\
\hline
95  & 40 & 19 & 0.1 \\
150 & 40 & 11 & 0.1 \\
\hline
\end{tabular}
\par\vspace{3pt}
\parbox{0.90\linewidth}{\footnotesize Notes: $\Delta_{\rm P}=w_{\rm P}^{-1/2}$ is the polarization map depth per frequency channel. Both channels observe the same sky area. The baseline multipole range is $30\leq\ell\leq1000$.}
\end{table}

We use $f_{\rm sky}=0.1$ and $30\leq\ell\leq1000$ for the baseline calculation. Here $f_{\rm sky}$ is the effective sky fraction entering the power-spectrum covariance. The lower multipole cut represents restricted access to large angular scales in ground-based observations, following the forecasting approach of \citet{Errard:2015cxa}. The upper cut is fixed throughout the parameter study. Sky coverage is described by its effective mode count in the present forecasts.

\subsection{Noise Model and Sensitivity Normalization}
\label{sec:noise}

For a frequency channel $\nu$, we write the beam-deconvolved polarization noise as
\begin{equation}
N_{\ell,\nu}^{\rm PP}=W_{\ell,\nu}
\left[1+\left(\frac{\ell}{\ell_{\rm knee}}\right)^{\alpha_\nu}\right],
\label{eq:noise}
\end{equation}
with the white-noise spectrum
\begin{equation}
W_{\ell,\nu}=(\Delta_{{\rm P},\nu}\,a)^2
\exp\left[\frac{\ell(\ell+1)(\theta_{{\rm FWHM},\nu}\,a)^2}{8\ln2}\right],
\label{eq:white}
\end{equation}
where $a=\pi/(180\times60)$ converts arcminutes to radians, and the numerical values of $\Delta_{{\rm P},\nu}$ and $\theta_{{\rm FWHM},\nu}$ are in $\mu\mathrm{K\,arcmin}$ and arcminutes, respectively. The second term in the brackets models noise that grows toward large angular scales, as in commonly used CMB survey forecasts \citep{SimonsObservatory:2018koc}. We take $\ell_{\rm knee}=30$, $\alpha_{95}=-2.5$, and $\alpha_{150}=-3.0$. The red and white contributions are equal at $\ell=\ell_{\rm knee}$ in each channel. Both receive the same beam-deconvolution factor, so the red term can turn upward at high $\ell$ even while its ratio to the white term decreases.

\begin{figure}[!htbp]
\centering
\includegraphics[width=\linewidth]{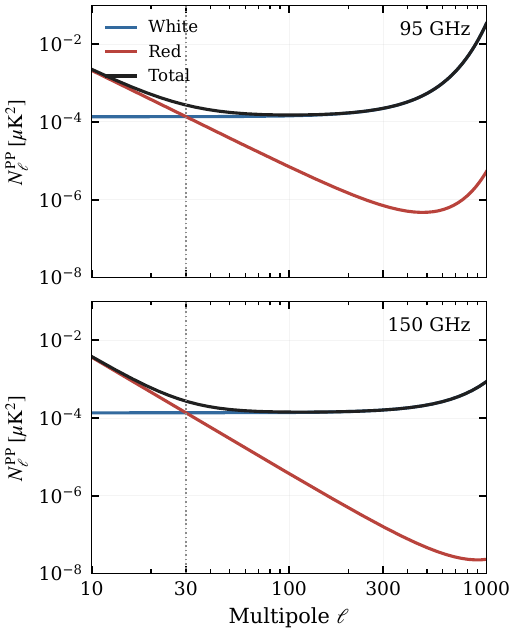}
\caption{Beam-deconvolved polarization noise in the two NHT-1 channels. Blue, red, and black curves show the white component, red component, and their sum. The vertical dotted line marks $\ell_{\rm knee}=30$, where the two components are equal. Curves below this multipole illustrate the noise model; the baseline likelihood begins at $\ell=30$}
\label{fig:rednoise}
\end{figure}

The noise in the frequency channels is assumed independent. Their full polarization noise spectra are combined with inverse-variance weights,
\begin{equation}
N_\ell^{\rm PP}=\left[\sum_\nu
\left(N_{\ell,\nu}^{\rm PP}\right)^{-1}\right]^{-1}.
\label{eq:multinoise}
\end{equation}
Thus the red component is included before combining channels. Setting it to zero gives the familiar white-noise combination \citep{1995PhRvD..52.4307K,Errard:2015cxa}. We use the same polarization noise for $EE$ and $BB$. Temperature noise contains only the white term and obeys
\begin{equation}
N_\ell^{\rm TT}=\frac12\left[\sum_\nu W_{\ell,\nu}^{-1}\right]^{-1},
\label{eq:tempnoise}
\end{equation}
corresponding to $\Delta_{{\rm T},\nu}=\Delta_{{\rm P},\nu}/\sqrt{2}$ for NHT-1. At low multipoles, before adding red noise, the two $40\,\mu\mathrm{K\,arcmin}$ channels have a combined white-noise depth of $40/\sqrt{2}\simeq28.3\,\mu\mathrm{K\,arcmin}$.

The overall noise amplitude is chosen by reference to the AliCPT pipeline forecasts of \citet{Ghosh:2022mje}. A per-channel depth of $\Delta_{\rm P}=40\,\mu\mathrm{K\,arcmin}$ gives $\sigma(r)=0.023$ with the likelihood below, close to the $\sim0.02$ sensitivity scale of those forecasts. This sets the reference sensitivity for the parameter study. The subsequent scans specify progressively lower map depths while holding the other model assumptions fixed, thereby evaluating the constraints associated with each assumed sensitivity.

Foreground separation remains part of the observational problem. The available methods include internal linear combinations, independent-component analysis, and spectral matching \citep{2004ApJ...612..633E,2003PhRvD..68l3523T,Maino:2001vz,Delabrouille:2002kz,Cardoso:2008qt}, parametric component separation \citep{2008ApJ...676...10E,stivoli2010maximum}, and map reconstruction or template-based approaches \citep{Martinez-Gonzalez:2003abe,Leach:2008fi,fernandez2012multiresolution}. Their performance for B-mode measurements has been explored with simulated polarization data \citep{Santos:2019csa,Remazeilles:2020rqw,Adak:2021lbu,Zhang:2020ltv,Carones:2022dnm}, including studies directed at AliCPT \citep{Ghosh:2022mje,Dou:2023vqg,Zhang:2024fg}. We adopt an idealized foreground-cleaned CMB model with zero residual-foreground power. The two bands specify its noise sensitivity. Under this assumption, the scans quantify how the constraint responds to map depth and mode coverage.

\subsection{Mock Likelihood and Parameter Inference}
\label{sec:likelihood}

\begin{figure}[!htbp]
\centering
\includegraphics[width=\linewidth]{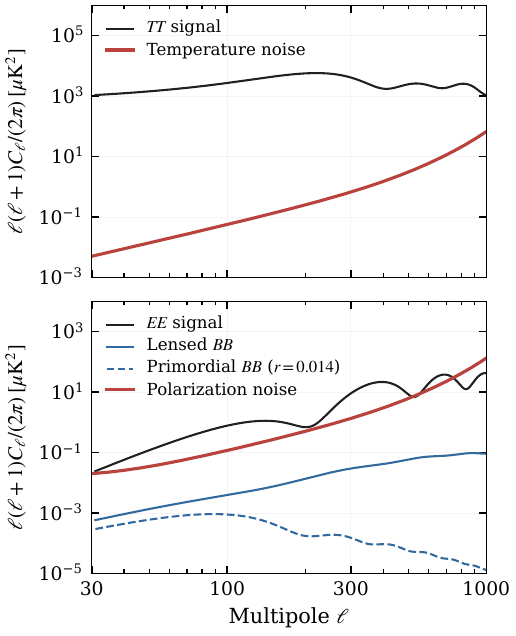}
\caption{Fiducial signal and noise spectra for NHT-1. The upper panel shows the temperature signal (black) and noise (red). The lower panel shows $EE$ (black), the total lensed $BB$ signal (solid blue), its primordial contribution for $r=0.014$ (dashed blue), and the full combined polarization noise (red). Both noise spectra use the channel combination in Section~\ref{sec:noise}}
\label{fig:spectra}
\end{figure}

We generate the mock CMB spectra in a $\Lambda$CDM+$r$ cosmology. The scalar-sector parameters are based on the \textit{Planck} 2018 best fit \citep{Planck:2018vyg} and are listed in Table~\ref{tab:fiducial}. We adopt $r=0.014$ at the pivot scale $k_*=0.05\,\mathrm{Mpc}^{-1}$, corresponding to the nonzero central value reported in the BICEP/\textit{Keck} 2018-season analysis \citep{BICEP:2021xfz}. This value defines the tensor amplitude of the fiducial mock sky. The $BB$ signal contains both the primordial and lensing contributions. Lensing B modes are retained throughout, with no delensing reduction. Figure~\ref{fig:spectra} compares the signal and noise spectra for the benchmark.

\begin{table}[!htbp]
\centering
\caption{Fiducial Cosmological Parameters}
\label{tab:fiducial}
\footnotesize
\begin{tabular*}{0.76\linewidth}{@{\extracolsep{\fill}}lc}
\hline
Parameter & Value \\
\hline
$\omega_{\rm b}$ & 0.02242 \\
$\omega_{\rm cdm}$ & 0.11933 \\
$100\theta_{\rm s}$ & 1.04101 \\
$\ln(10^{10}A_{\rm s})$ & 3.047 \\
$n_{\rm s}$ & 0.9665 \\
$\tau_{\rm reio}$ & 0.0561 \\
$r$ & 0.014 \\
\hline
\end{tabular*}
\par\vspace{3pt}
\parbox{0.90\linewidth}{\footnotesize Notes: The pivot scale is $k_*=0.05\,\mathrm{Mpc}^{-1}$.}
\end{table}

For NHT-1, we use the mock-CMB likelihood implemented in \textsc{MontePython}, including the $TT$, $TE$, $EE$, and $BB$ spectra. In the $(T,E,B)$ basis, the signal-plus-noise covariance of the harmonic coefficients is
\begingroup
\setlength{\arraycolsep}{1pt}
\begin{equation}
\mathbf M_\ell=\begin{pmatrix}
C_\ell^{TT}+N_\ell^{TT}&C_\ell^{TE}&0\\
C_\ell^{TE}&C_\ell^{EE}+N_\ell^{\rm PP}&0\\
0&0&C_\ell^{BB}+N_\ell^{\rm PP}
\end{pmatrix},
\label{eq:covariance}
\end{equation}
\endgroup
where the theoretical spectra depend on the sampled cosmological parameters $\boldsymbol{\theta}$. Temperature and polarization noise are assumed uncorrelated, $N_\ell^{TE}=0$, while the cosmological cross-spectrum $C_\ell^{TE}$ is retained. The $TB$ and $EB$ spectra vanish in the adopted cosmology. The $BB$ spectrum includes the primordial and lensing contributions, and the polarization noise is given by Equation~\eqref{eq:multinoise}.

Let $\widehat{\mathbf M}_\ell$ denote the corresponding signal-plus-noise matrix evaluated at the fiducial cosmology. The likelihood is
\begin{equation}
\begin{split}
-2\ln\mathcal L_{\rm N}
={}&\sum_{\ell=\ell_{\min}}^{\ell_{\max}}(2\ell+1)f_{\rm sky}\biggl[
\operatorname{Tr}\!\left(\widehat{\mathbf M}_\ell\mathbf M_\ell^{-1}\right)\\
&\qquad+\ln\frac{\det\mathbf M_\ell}{\det\widehat{\mathbf M}_\ell}-3\biggr].
\end{split}
\label{eq:likelihood}
\end{equation}
The covariance and its determinant vary with the sampled cosmology. This expression uses the Gaussian harmonic-mode approximation, with $f_{\rm sky}$ specifying the effective number of modes. Correlations between multipoles induced by a specific sky mask are neglected \citep{Hamimeche:2008ai}.

The theoretical spectra are computed with \textsc{class} \citep{blas2011cosmic}, and the posterior is sampled with \textsc{MontePython} using the Metropolis--Hastings algorithm \citep{audren2013conservative,2019PDU....24..260B}. We vary the parameters of the $\Lambda$CDM+$r$ model and run the chains until the Gelman--Rubin statistic satisfies $R-1\sim0.001$ for each fundamental parameter \citep{Gelman:1992zz}. For an asymmetric marginalized constraint written as ${r_{\rm c}}^{+\delta r_+}_{-\delta r_-}$, we summarize its uncertainty by
\begin{equation}
\sigma(r)\equiv\frac{\delta r_++\delta r_-}{2}.
\label{eq:sigma}
\end{equation}
We use this interval-width convention throughout the sensitivity comparisons and retain the asymmetric errors in the posterior plots. It accommodates the physical boundary at $r=0$; for a skewed distribution it can differ from the posterior standard deviation.

\subsection{Joint Forecasts with Planck-like Data}
\label{sec:joint}

We combine NHT-1 with the \texttt{fake\_planck\_realistic} mock likelihood in \textsc{MontePython}, using the noise prescription of \citet{Brinckmann:2018owf}. This is a synthetic Planck-like data set with the same fiducial cosmology as NHT-1. Its temperature and polarization noise are read from the public file \texttt{noise\_fake\_planck\_realistic\_two.dat}. The file also supplies the lensing-reconstruction noise. The associated configuration uses the 100 and 143\,GHz channels and omits 217\,GHz. Table~\ref{tab:joint} gives the nominal channel parameters; the tabulated noise spectra define the numerical input.

\begin{table*}[!htbp]
\centering
\caption{Parameters of the Joint Mock Data}
\label{tab:joint}
\small
\begin{tabular}{cccc}
\hline
Frequency (GHz) & $\theta_{\rm FWHM}$ (arcmin) &
$\Delta_{\rm T}$ ($\mu$K arcmin) & $\Delta_{\rm P}$ ($\mu$K arcmin) \\
\hline
\multicolumn{4}{c}{Low-$\ell$ Planck-like data: $2\leq\ell\leq29$, $f_{\rm sky}=0.57$} \\
100 & 10.0 & 68.0 & 109.0 \\
143 & 7.1 & 42.6 & 80.94 \\
\hline
\multicolumn{4}{c}{Mid-$\ell$ Planck-like data: $30\leq\ell\leq1000$, $f_{\rm sky}=0.47$} \\
100 & 10.0 & 68.0 & 109.0 \\
143 & 7.1 & 42.6 & 80.94 \\
\hline
\multicolumn{4}{c}{NHT-1: $30\leq\ell\leq1000$, $f_{\rm sky}=0.1$} \\
95 & 19.0 & 28.3 & 40.0 \\
150 & 11.0 & 28.3 & 40.0 \\
\hline
\multicolumn{4}{c}{High-$\ell$ Planck-like data: $1001\leq\ell\leq3000$, $f_{\rm sky}=0.57$} \\
100 & 10.0 & 68.0 & 109.0 \\
143 & 7.1 & 42.6 & 80.94 \\
\hline
\end{tabular}
\par\vspace{3pt}
\parbox{0.90\linewidth}{\footnotesize Notes: In the shared multipole interval, the NHT-1 and Planck-like contributions occupy separate sky fractions of 0.1 and 0.47. The Planck-like channel depths follow the public configuration comments: pixel sensitivities of $(6.8,6.0)\,\mu$K in temperature and $(10.9,11.4)\,\mu$K in polarization, multiplied by beam widths of $(10.0,7.1)$ arcmin. The tabulated noise file is used directly. The NHT-1 temperature depth follows Equation~\eqref{eq:tempnoise}, rounded to one decimal place.}
\end{table*}

The data are divided into the multipole and sky-coverage ranges listed in Table~\ref{tab:joint}. Planck-like data cover $2\leq\ell\leq29$ and $1001\leq\ell\leq3000$ with $f_{\rm sky}=0.57$. Within $30\leq\ell\leq1000$, we use NHT-1 spectra over $f_{\rm sky}=0.1$ and Planck-like spectra over the remaining $f_{\rm sky}=0.47$. The NHT-1 contribution thus replaces the Planck-like contribution on the same area in this interval. The likelihood combines the separate ranges with their respective noise levels and sky fractions.

The Planck-like temperature and polarization noise are independent inputs; the NHT-1 relation in Equation~\eqref{eq:tempnoise} is not imposed on them. At low multipoles the tabulated noise corresponds approximately to combined depths of $36.1\,\mu\mathrm{K\,arcmin}$ in temperature and $65.0\,\mu\mathrm{K\,arcmin}$ in polarization. The disjoint sky fractions avoid counting the same modes twice within the effective-area approximation.

The Planck-like component uses the same mock-CMB likelihood form for temperature, E-mode polarization, and lensing reconstruction, with direct $BB$ information disabled and no delensing. Defining the deflection field by $d_{\ell m}=\sqrt{\ell(\ell+1)}\,\phi_{\ell m}$, its covariance matrix in the $(T,E,d)$ basis is
\begin{equation}
\mathbf S_\ell=\begin{pmatrix}
C_\ell^{TT}+N_\ell^{TT}&C_\ell^{TE}&0\\
C_\ell^{TE}&C_\ell^{EE}+N_\ell^{\rm PP}&0\\
0&0&C_\ell^{dd}+N_\ell^{dd}
\end{pmatrix},
\label{eq:planckcov}
\end{equation}
The off-diagonal entries retain the cosmological $TE$ signal. Temperature and polarization noise are assumed uncorrelated, $N_\ell^{TE}=0$, following \citet{Brinckmann:2018owf}. The temperature--deflection cross-correlation is neglected, as in the public configuration.

The Planck-like log likelihood follows Equation~\eqref{eq:likelihood} with $\mathbf M_\ell$ and $\widehat{\mathbf M}_\ell$ replaced by $\mathbf S_\ell$ and $\widehat{\mathbf S}_\ell$, respectively, where $\widehat{\mathbf S}_\ell$ is evaluated at the fiducial cosmology. We evaluate it separately in each Planck-like range in Table~\ref{tab:joint}, using the corresponding noise spectra, multipole limits, and sky fraction, and add the NHT-1 log likelihood. Appendix~\ref{app:spectra} shows the temperature, polarization, and Planck-like lensing-reconstruction noise. In Table~\ref{tab:depth}, ``NHT-1 + Planck'' denotes this joint mock-data configuration.

\section{Forecasts and Parameter Dependence}
\label{sec:results}

\subsection{Benchmark and Joint Constraints}
\label{sec:baseline}

\begin{figure}[!htbp]
\centering
\includegraphics[width=\linewidth]{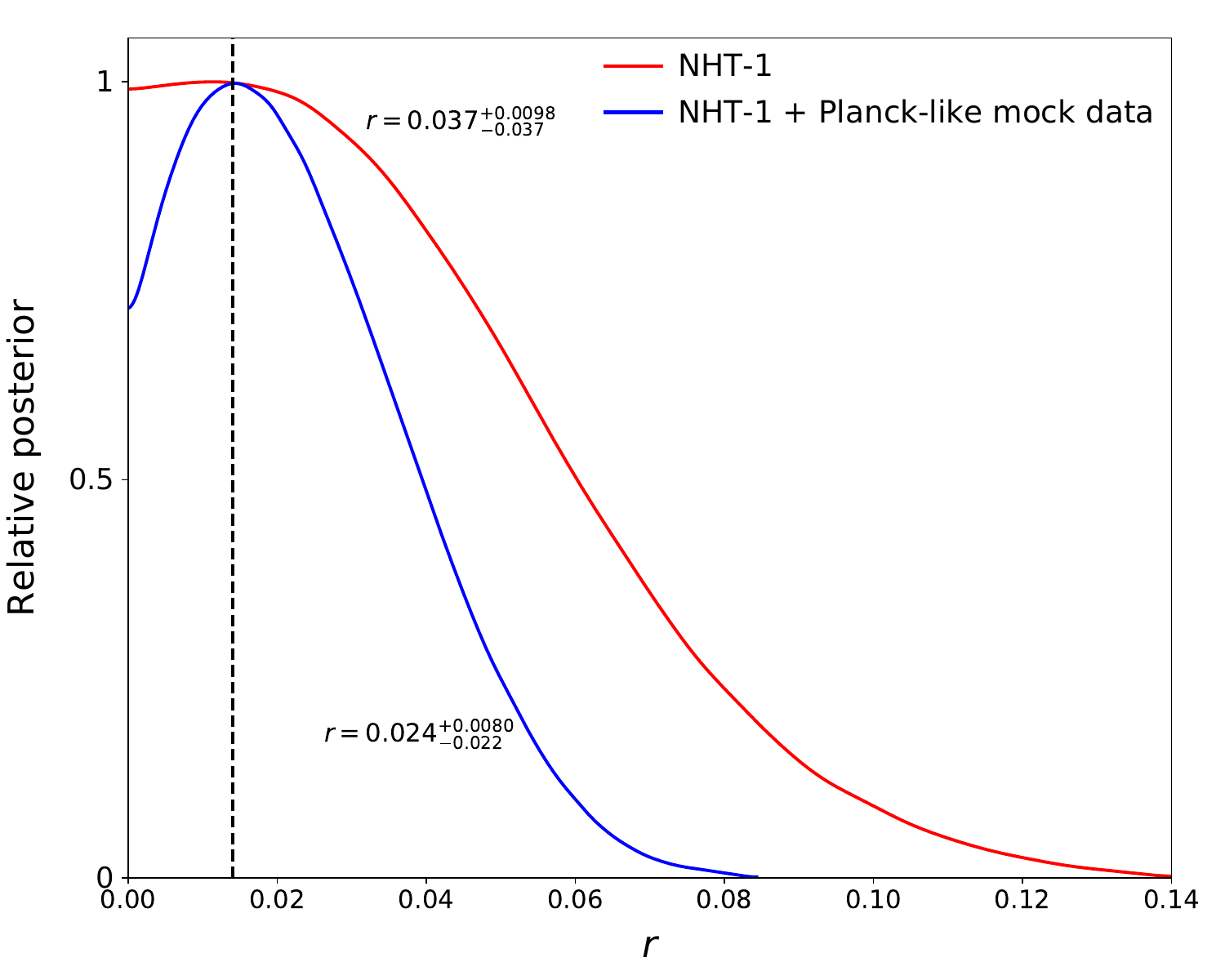}
\caption{Marginalized posterior distributions of $r$ for NHT-1 (red) and NHT-1 combined with Planck-like mock data (blue). The red curve sets the unit peak height on the vertical axis. The dashed vertical line marks the fiducial value $r=0.014$. The annotations give the central values and asymmetric marginalized errors}
\label{fig:posterior}
\end{figure}

Figure~\ref{fig:posterior} shows the marginalized $r$ distributions for NHT-1 and the joint configuration. The benchmark gives $\sigma(r)=0.023$, setting the reference point for the parameter variations below. Adding Planck-like information reduces this uncertainty to 0.015. The asymmetric intervals shown in the figure are summarized using Equation~\eqref{eq:sigma}.

The joint configuration supplies complementary temperature, E-mode polarization, and lensing information across a broader range of scales and sky coverage. For observational context, the \textit{Planck} 2018 inflation analysis obtained $r_{0.002}<0.10$ at 95\% confidence from $TT,TE,EE+\mathrm{lowE}+\mathrm{lensing}$ \citep{Planck:2018jri}. That upper limit is quoted at $k=0.002\,\mathrm{Mpc}^{-1}$, while the forecasts here report interval widths at $k_*=0.05\,\mathrm{Mpc}^{-1}$. Within the common mock-data model, the comparison in Figure~\ref{fig:posterior} directly quantifies the benefit of combining northern polarization measurements with Planck-like information.

\subsection{Dependence on Map Depth}
\label{sec:depth}

We vary the polarization map depth in both frequency channels together, taking $\Delta_{\rm P}=40$, 30, 20, and $10\,\mu\mathrm{K\,arcmin}$. The beam widths, sky fraction, multipole range, and dimensionless red-noise parameters remain fixed. The red and white components have equal amplitudes at $\ell_{\rm knee}$; this scan changes their common normalization while preserving the shape of the noise spectrum. Table~\ref{tab:depth} lists the resulting marginalized errors on $r$ and $n_{\rm s}$.

\begin{table*}[!htbp]
\centering
\caption{Constraints as a Function of Map Depth}
\label{tab:depth}
\small
\setlength{\tabcolsep}{8pt}
\begin{tabular}{lcccc}
\hline
$\Delta_{\rm P}$ ($\mu$K arcmin) & 40 & 30 & 20 & 10 \\
\hline
\multicolumn{5}{c}{NHT-1} \\
$\sigma(r)$ & 0.023 & 0.015 & 0.0092 & 0.0031 \\
$\sigma(n_{\rm s})$ & 0.012 & 0.011 & 0.010 & 0.0096 \\
\hline
\multicolumn{5}{c}{NHT-1 + Planck} \\
$\sigma(r)$ & 0.015 & 0.013 & 0.0088 & 0.0030 \\
$\sigma(n_{\rm s})$ & 0.0040 & 0.0040 & 0.0039 & 0.0038 \\
\hline
\end{tabular}
\par\vspace{3pt}
\parbox{0.90\linewidth}{\footnotesize Notes: The same map depth is assigned to each NHT-1 channel. The joint forecasts use the Planck-like configuration in Table~\ref{tab:joint}. All other survey parameters are held fixed.}
\end{table*}

The constraint on $r$ improves strongly as the maps become deeper. For NHT-1 alone, the error decreases from 0.023 to 0.0031 across the scan. The corresponding joint errors decrease from 0.015 to 0.0030. At $40\,\mu\mathrm{K\,arcmin}$, Planck-like data provide a substantial improvement over NHT-1 alone. At $10\,\mu\mathrm{K\,arcmin}$, the two $r$ constraints are nearly equal. Thus deeper northern-sky polarization data increasingly determine the precision on $r$ in this joint configuration. The scan connects the benchmark sensitivity near $0.02$ to a prospective precision of a few $10^{-3}$ under the adopted foreground and lensing assumptions.

The behavior of $n_{\rm s}$ is different. Its NHT-1-only error changes from 0.012 to 0.0096, whereas the joint error remains close to 0.004 throughout the scan. The Planck-like temperature and polarization information therefore continues to be important for the scalar spectral index even when the northern-sky measurement dominates the improvement in $r$.

\subsection{Access to Large Angular Scales}
\label{sec:largescales}

\begin{figure}[!htbp]
\centering
\includegraphics[width=\linewidth]{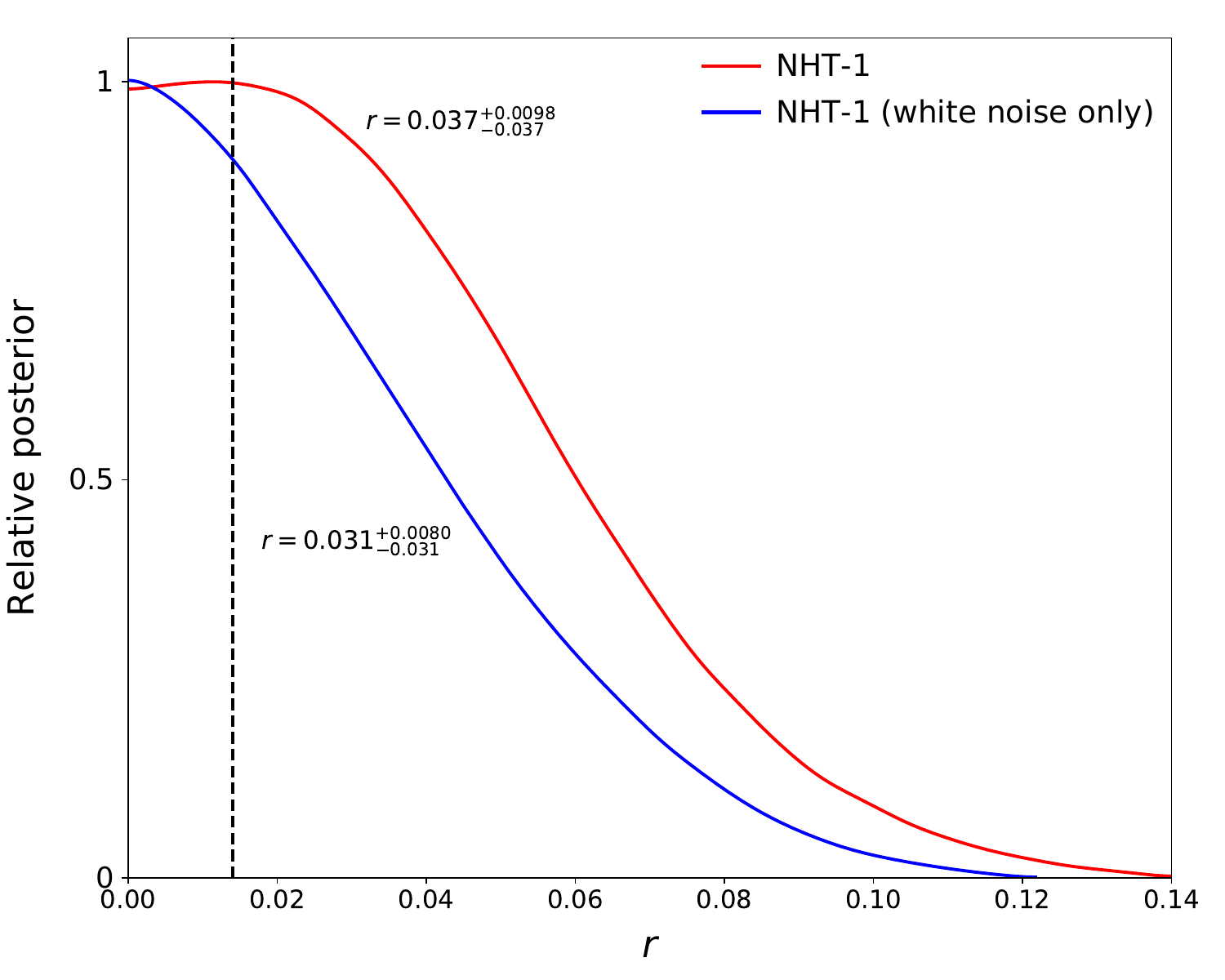}
\caption{Marginalized posterior distributions of $r$ for NHT-1 with the baseline red-noise component (red) and with white noise only (blue). The red curve sets the unit peak height on the vertical axis. The white-noise normalization is the same in both cases. The dashed vertical line marks $r=0.014$}
\label{fig:redposterior}
\end{figure}

We examine two aspects of large-angular-scale sensitivity: the red-noise contribution within the observed multipole range, and the loss of modes when the lower multipole cut is raised. Figure~\ref{fig:redposterior} compares the baseline forecast with a calculation in which the red term in Equation~\eqref{eq:noise} is removed. With the same white-noise depth, the error decreases from $\sigma(r)=0.023$ to 0.019. This comparison quantifies the effect of the adopted red-noise amplitude and slope over $30\leq\ell\leq1000$.

\begin{figure}[!htbp]
\centering
\includegraphics[width=\linewidth]{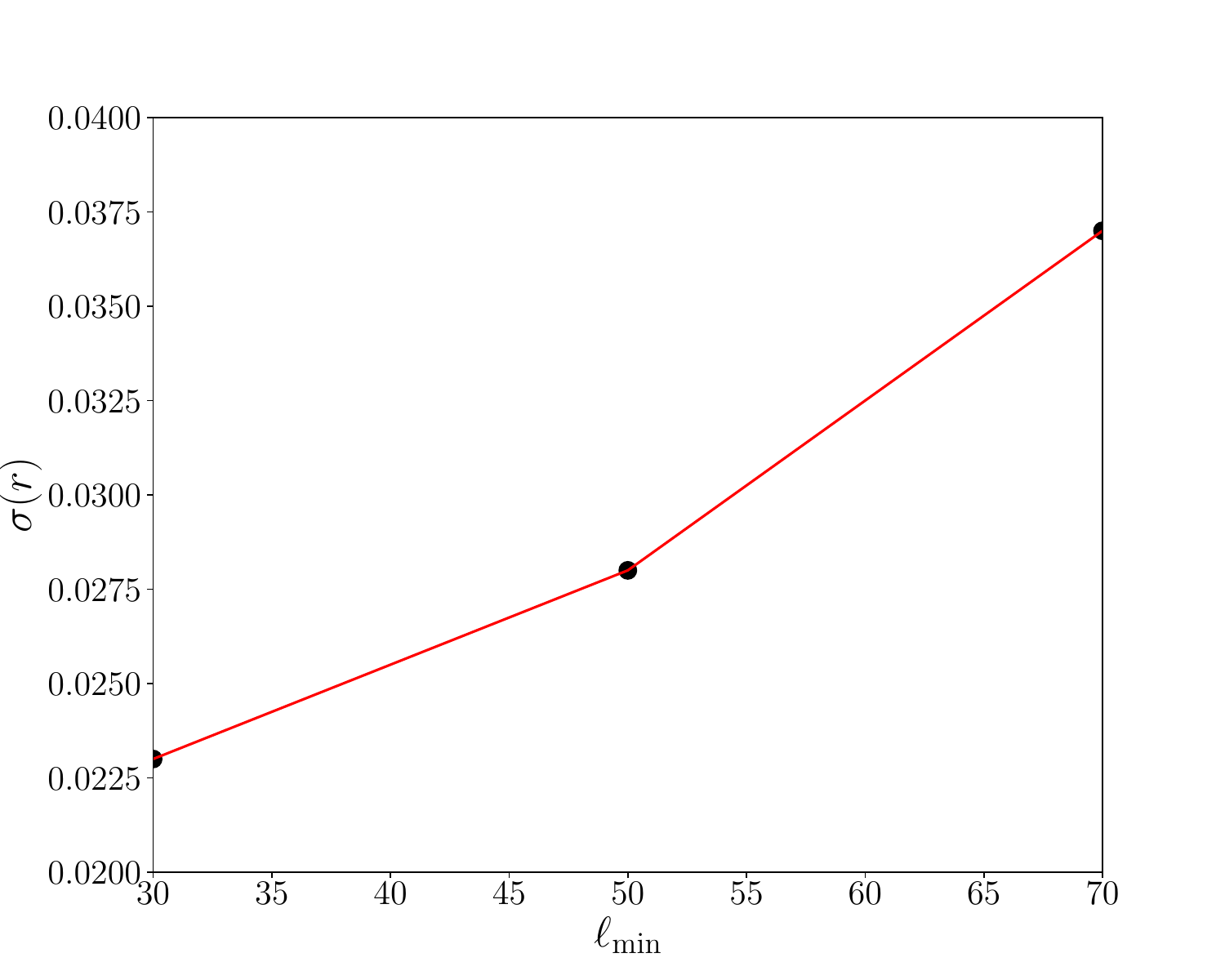}
\caption{Dependence of the NHT-1 constraint on the minimum multipole. The upper multipole cut, map depth, sky fraction, and noise-model parameters are fixed. Black points show the forecasts, joined by the red line}
\label{fig:lmin}
\end{figure}

Figure~\ref{fig:lmin} shows the effect of increasing $\ell_{\min}$ from 30 to 50 and 70 while retaining the baseline noise model. The error grows from 0.023 to approximately 0.028 and 0.037, respectively. The modes lost in this scan contain part of the large-scale primordial B-mode signal. Retaining these modes improves sensitivity even at unchanged map depth.

\subsection{Dependence on Survey Area at Fixed Map Depth}
\label{sec:area}

\begin{figure}[!htbp]
\centering
\includegraphics[width=\linewidth]{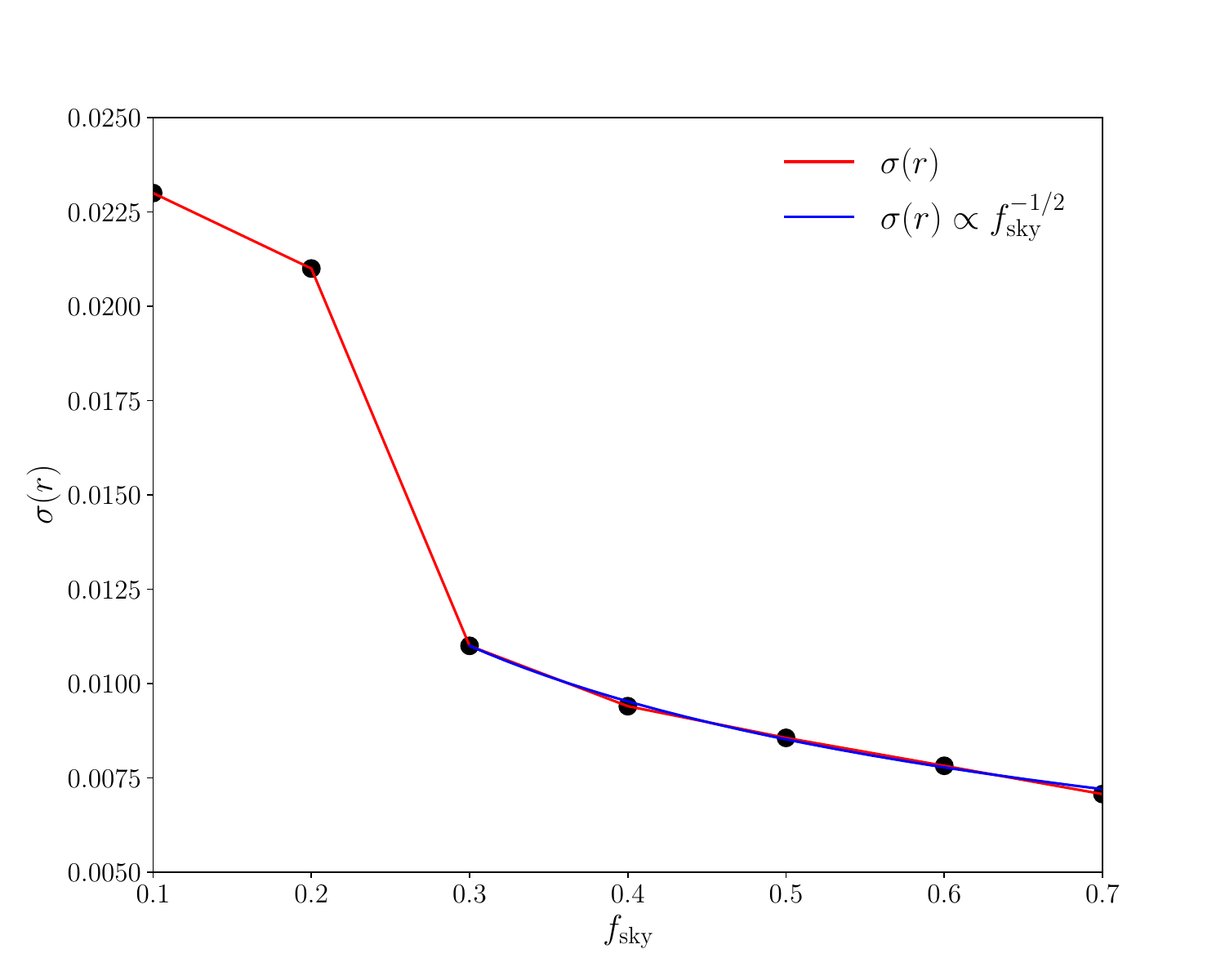}
\caption{Dependence of $\sigma(r)$ on survey area at fixed map depth and fixed noise-model parameters. Black points and the red connecting line show the MCMC forecasts. The blue curve is a reference proportional to $f_{\rm sky}^{-1/2}$, normalized at $f_{\rm sky}=0.3$}
\label{fig:fsky}
\end{figure}

Finally, we vary $f_{\rm sky}$ from 0.1 to 0.7 while holding the map depth and other noise parameters fixed. Figure~\ref{fig:fsky} shows that $\sigma(r)$ decreases from 0.023 to 0.0071 over this range. Here increasing the area increases the number of measured modes at the same assumed noise level per unit solid angle.

For comparison, the figure includes the Gaussian mode-counting reference $\sigma(r)\propto f_{\rm sky}^{-1/2}$, normalized at $f_{\rm sky}=0.3$. The connecting line joins the sampled MCMC forecasts.

The larger-area cases describe an extended effective footprint with uniform depth. They provide an idealized reference for combinations of northern, southern, and space-based observations, with overlap counted once.

\FloatBarrier
\section{Implications for Northern-sky CMB Observations}
\label{sec:discussion}

The forecasts show the scope for improving tensor constraints from northern polarization observations as their map depth increases. At $f_{\rm sky}=0.1$, the NHT-1-only uncertainty decreases from $0.023$ to $0.0031$ across the depth scan. This connects an AliCPT-motivated benchmark to a level at which northern maps could make a substantial contribution to the search for primordial B modes. Each point specifies the cosmological return associated with an assumed survey sensitivity. The study therefore supplies a quantitative reference for assessing future northern observations as their sensitivity develops.

Access to large angular scales is an additional determinant of that return. Raising $\ell_{\min}$ from 30 to 70 increases the baseline uncertainty from $0.023$ to $0.037$ without changing the map depth. The comparison identifies the information carried by those modes and motivates preserving it in future observing and analysis strategies. The area scan addresses a further contribution: more CMB modes become available when a larger region is measured at the same depth. Depth, angular-scale access, and effective area thus describe complementary aspects of the sensitivity explored here.

The Planck-like combination places these results in the context of complementary CMB observations. At the benchmark depth, external information improves the uncertainty on $r$ from $0.023$ to $0.015$. At $10\,\mu\mathrm{K\,arcmin}$ per channel, the northern-only and joint uncertainties are $0.0031$ and $0.0030$, respectively, while the external data continue to sharpen $n_{\rm s}$. The changing balance indicates that sufficiently deep northern polarization maps can supply much of the tensor sensitivity in this combination. Broader temperature and E-mode measurements remain valuable for the wider cosmological parameter set.

The geographical motivation extends to joint ground-based observations. AliCPT design studies describe complementary sky access with southern experiments, including BICEP and the Simons Array \citep{Li:2017drr,2018NatAs...2..104L}. Combining deep observations of additional northern and southern low-foreground fields can enlarge the measured mode set. In common fields, cross-spectra and comparisons between independent instruments can test signal consistency. The northern sensitivity benchmarks established here motivate a subsequent quantitative analysis of these possibilities using the actual survey footprints, frequency coverage, depth variations, and overlap.

Angular resolution provides another connection between surveys. Degree-scale polarization measurements probe the recombination B-mode signal, while higher-resolution maps support lensing reconstruction \citep{2003PhRvD..67h3002O} and delensing \citep{Errard:2015cxa,CMB-S4:2020lpa}. AliCPT studies of lensing reconstruction and neutrino cosmology illustrate the wider information available from northern CMB measurements \citep{Han:2023gvr,Zhang:2021ecp}. A combined analysis can connect these uses of different angular scales.

A natural next step is to evaluate northern and southern observations in a common likelihood, with residual foregrounds, survey-specific noise and filtering, and instrumental systematics. Existing AliCPT foreground studies and work on polarization calibration provide starting points for those extensions \citep{Zhang:2024fg,LiteBIRD:2021hlz}. The present idealized forecasts establish the dependence on survey sensitivity and mode coverage that such analyses can build upon. They identify the potential for northern observations to contribute both tighter tensor constraints and independent checks within an international CMB program.

\section{Conclusions}
\label{sec:conclusions}

We have assessed the potential of northern-sky CMB polarization observations through an idealized survey model motivated by AliCPT. The forecasts retain lensing B modes and assume negligible residual foreground power. For the two-band benchmark at 95 and 150\,GHz, with a polarization map depth of $40\,\mu\mathrm{K\,arcmin}$ per channel and $f_{\rm sky}=0.1$, the adopted $\Lambda$CDM+$r$ cosmology with $r=0.014$ gives $\sigma(r)=0.023$. Combining the benchmark with Planck-like mock data improves this to $0.015$.

The parameter scans quantify the improvement associated with deeper maps, access to large angular scales, and greater effective area. At a per-channel depth of $10\,\mu\mathrm{K\,arcmin}$, the northern-only and joint uncertainties reach $0.0031$ and $0.0030$. Removing the adopted red-noise term gives $\sigma(r)=0.019$ at the baseline white-noise depth, whereas raising $\ell_{\min}$ from 30 to 70 increases the error to $0.037$. Extending sky coverage improves the constraint at fixed depth. As the northern maps deepen, they increasingly determine the precision on $r$ in the joint forecast, while Planck-like information remains important for $n_{\rm s}$.

These results connect an AliCPT-motivated sensitivity benchmark to prospective tensor constraints at the level of a few $10^{-3}$. They provide a first stage in assessing the contribution of improved northern polarization measurements to a global CMB program. Detailed combinations with southern surveys, incorporating their masks, frequency coverage, noise, foregrounds, and instrumental effects, can use these benchmarks to evaluate the precision and consistency tests available from joint observations.

\backmatter
\begingroup
\small

\section*{Acknowledgements}
We are grateful to Hong Li, Siyu Li, Yang Liu, Wentao Luo, Toshiya Namikawa, Larissa Santos, and Rui Shi for valuable discussions.

We acknowledge the use of computing facilities of the astronomy department and the LINDA \& JUDY clusters of the particle cosmology group at USTC. Generative large language models assisted with manuscript editing. The authors are responsible for the scientific content.

\section*{Statements and Declarations}

\subsection*{Funding}
This work is supported in part by the National Key Research and Development Program of China (2021YFC2203100), the Youth Project of Scientific Research Program of Anhui Provincial Depart ment of Education (2025AHGXZK40234), the Hefei Institute of Technology Talent Research Fund (2025KY41), Young Interdisciplinary Innovation Team of the Chinese Academy of Sciences (JCTD-2022-20), National Natural Science Foundation of China (12261131497, 11653002, 12003029), the 111 Project for ``Observational and Theoretical Research on Dark Matter and Dark Energy'' (B23042), Fundamental Research Funds for Central Universities, China Scholarship Council Innovation Talent Funds, University of Science and Technology of China Fellowship for International Cooperation, University of Science and Technology of China Research Funds of the Double First-Class Initiative, and the Chinese Academy of Sciences project for young scientists in basic research (YSBR-006).

\subsection*{Data availability}
The survey specifications, fiducial cosmology, and numerical forecast results are presented in this article. Additional numerical data supporting the results are available from the corresponding authors upon reasonable request.

\subsection*{Code availability}
The calculations use the publicly available \textsc{class} and \textsc{MontePython} packages \citep{blas2011cosmic,audren2013conservative,2019PDU....24..260B}. Analysis configurations and study-specific code needed to reproduce the results are available from the corresponding authors upon reasonable request.

\subsection*{Competing interests}
The authors declare no competing interests.
\endgroup

\clearpage
\onecolumn
\begin{appendices}
\setcounter{figure}{6}
\renewcommand{\thefigure}{\arabic{figure}}
\section{Signal and Noise Spectra for the Joint Forecast}
\label{app:spectra}

\begin{figure}[!htbp]
\centering
\includegraphics[width=0.94\textwidth]{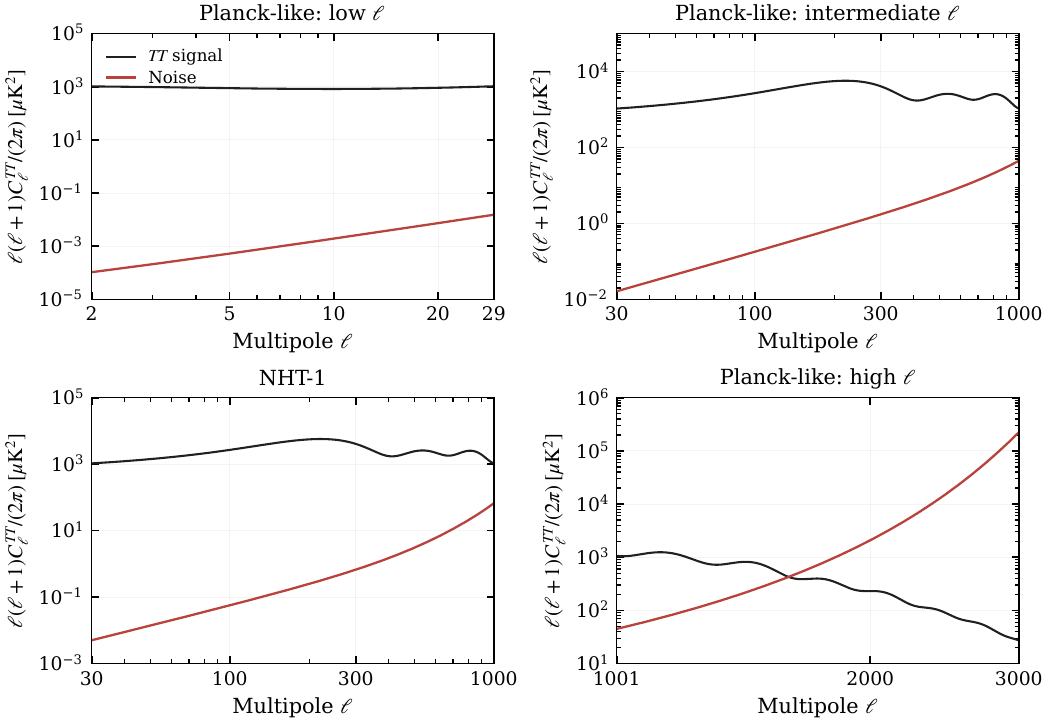}
\caption{Temperature signal and noise spectra for the joint configuration. The panels show low-$\ell$ Planck-like data (upper left), mid-$\ell$ Planck-like data (upper right), NHT-1 (lower left), and high-$\ell$ Planck-like data (lower right). Black curves show the fiducial $TT$ signal and red curves show the noise. The sky fractions and multipole ranges are listed in Table~\ref{tab:joint}}
\label{fig:jointtt}
\end{figure}

\begin{figure}[!htbp]
\centering
\includegraphics[width=0.94\textwidth]{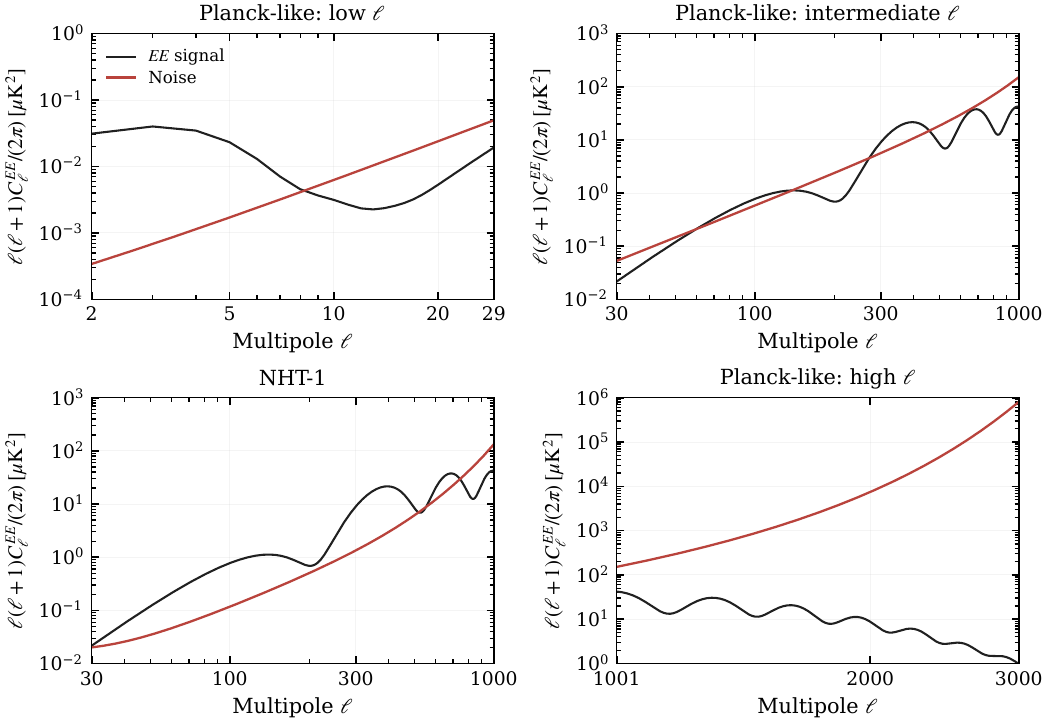}
\caption{E-mode polarization signal (black) and noise (red) for the four parts of the joint mock-data configuration. The panel ordering is the same as in Figure~\ref{fig:jointtt}}
\label{fig:jointee}
\end{figure}

Figures~\ref{fig:jointtt} and \ref{fig:jointee} display the temperature and E-mode polarization spectra for the four data ranges in Table~\ref{tab:joint}. The panels distinguish the low- and high-multipole Planck-like contributions from the NHT-1 and Planck-like contributions in the shared multipole interval. Their sky fractions enter the covariance as specified in Section~\ref{sec:joint}.

\begin{figure}[!htbp]
\centering
\includegraphics[width=0.94\textwidth]{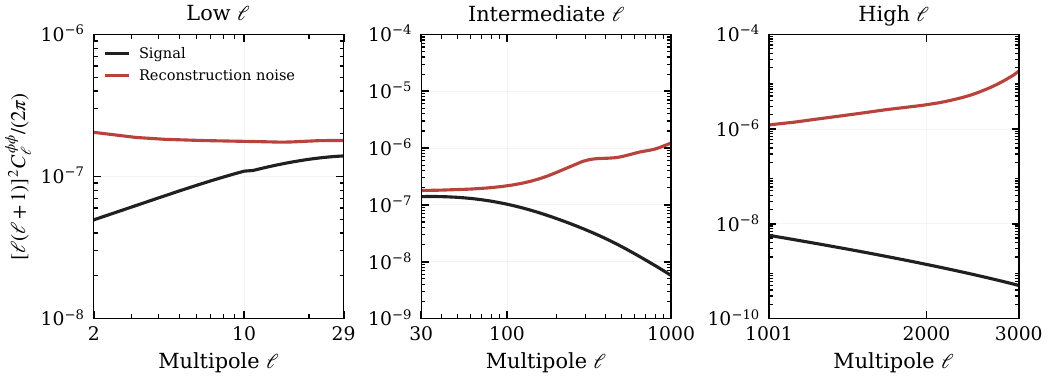}
\caption{Lensing-potential signal (black) and reconstruction noise (red) for the low-, intermediate-, and high-multipole Planck-like components. The reconstruction noise is read directly from the same public file as the temperature and polarization noise. The sky fractions enter the likelihood through the mode-count factor}
\label{fig:jointphi}
\end{figure}

The NHT-1 noise curves follow Equations~\eqref{eq:noise}--\eqref{eq:tempnoise}; the Planck-like curves use the public tabulated noise. Figure~\ref{fig:jointphi} shows the lensing-potential signal and reconstruction noise in the three Planck-like multipole ranges. The fourth column of the public noise file is $[\ell(\ell+1)]^2N_\ell^{\phi\phi}/(2\pi)$, which is converted to $N_\ell^{dd}$ in the likelihood. The NHT-1 $BB$ signal and noise are shown in Figure~\ref{fig:spectra}.

\end{appendices}
\clearpage
\twocolumn
\renewcommand{\bibfont}{\small}
\setlength{\bibsep}{0.7pt}
\bibliographystyle{epjc-num}
\bibliography{references}

\end{document}